\documentclass[letterpaper]{article} 
\usepackage{aaai24}  
\usepackage{times}  
\usepackage{helvet}  
\usepackage{courier}  
\usepackage[hyphens]{url}  
\usepackage{graphicx} 
\usepackage{natbib}  
\usepackage{caption} 
\usepackage{algorithm}
\usepackage[noend]{algorithmic}
\usepackage{amsmath}
\usepackage{color}
\usepackage{bm}

\usepackage{multirow, array, xcolor}
\newcolumntype{A}{>{\raggedright\arraybackslash}p{10cm}}
\newcolumntype{B}{>{\centering\arraybackslash}p{2.5cm}}

\usepackage{newfloat}
\usepackage{listings}
\DeclareCaptionStyle{ruled}{labelfont=normalfont,labelsep=colon,strut=off} 
\floatstyle{ruled}
\newfloat{listing}{tb}{lst}{}
\floatname{listing}{Listing}
\title{Quantifying gendered citation imbalance in computer science conferences}

\author{
    Kazuki Nakajima\textsuperscript{\rm 1},
    Yuya Sasaki\textsuperscript{\rm 2},
    Sohei Tokuno\textsuperscript{\rm 3},
    George Fletcher\textsuperscript{\rm 4}
}

\affiliations{
    \textsuperscript{\rm 1}Tokyo Metropolitan University\\
    \textsuperscript{\rm 2}Osaka University\\
    \textsuperscript{\rm 3}Nara Institute of Science and Technology\\
    \textsuperscript{\rm 4}Eindhoven University of Technology\\
    nakajima@tmu.ac.jp, sasaki@ist.osaka-u.ac.jp, tokuno.sohei@cdg.naist.jp, g.h.l.fletcher@tue.nl
}

\usepackage{bibentry}

\begin{document}

\maketitle


\begin{abstract}
The number of citations received by papers often exhibits imbalances in terms of author attributes such as country of affiliation and gender. 
While recent studies have quantified citation imbalance in terms of the authors' gender in journal papers, the computer science discipline, where researchers frequently present their work at conferences, may exhibit unique patterns in gendered citation imbalance. 
Additionally, understanding how network properties in citations influence citation imbalances remains challenging due to a lack of suitable reference models.
In this paper, we develop a family of reference models for citation networks and investigate gender imbalance in citations between papers published in computer science conferences.
By deploying these reference models, we found that homophily in citations is strongly associated with gendered citation imbalance in computer science, whereas heterogeneity in the number of citations received per paper has a relatively minor association with it.
Furthermore, we found that the gendered citation imbalance is most pronounced in papers published in the highest-ranked conferences, is present across different subfields, and extends to citation-based rankings of papers.
Our study provides a framework for investigating associations between network properties and citation imbalances, aiming to enhance our understanding of the structure and dynamics of citations between research publications.
\end{abstract}

\section{Introduction}

Citation analysis serves as a standard bibliometric tool for assessing the impact of different papers, researchers, institutions, and countries \cite{waltman2016, zeng2017, fortunato2018}.
The outcomes of such analyses are pivotal, as they influence various facets of research evaluation, including the hiring and promotion of researchers \cite{moed2006, moher2018}.
However, achieving unbiased citation analysis remains a challenge due to the diverse citation practices adopted by individual researchers and across research disciplines \cite{radicchi2008}.
Consequently, citation imbalances persist, such as the over-citation of certain journal groups \cite{kojaku2021} and country groups \cite{gomez2022}.

\noindent
{\bf Motivation.} 
Gender imbalance persists in computer science across various aspects, including educational attainment \cite{oecd}, faculty representation \cite{laberge2022}, and career progression \cite{lietz2024}.
Addressing this imbalance is crucial for promoting gender equality in education, academia, industry, and society \cite{jaccheri2020}.
While recent studies have highlighted citation imbalances related to authors' gender in neuroscience and physics \cite{dworkin2020, teich2022}, the extent of gendered citation imbalance in computer science remains largely unknown.
Previous findings from other disciplines may not be directly applied to the computer science discipline due to unique publication practices.
Unlike many disciplines where journals are the primary publication venue, computer scientists often choose conferences to present their work \cite{freyne2010, vrettas2015}.
In addition, conference ranks, rather than journal impact factors, may influence citation practices of computer scientists \cite{freyne2010, vrettas2015}. 
These distinct research practices, combined with the technical challenges of inferring authors' gender, make it difficult to quantify gendered citation imbalances in computer science using existing data sources alone.

We represent citations between papers as a directed network to quantify gendered citation imbalance.
Citation networks are valuable for understanding the scientific knowledge space, as they exhibit various structural and dynamical properties \cite{zeng2017}.
Notably, citation networks often exhibit {\it homophily in citations}, where papers tend to cite others with similar characteristics \cite{ciotti2016}. 
Moreover, there exists {\it heterogeneity in the number of citations received per paper}, with most papers garnering few citations while a small fraction receives a substantial number \cite{eom2011}.
Given that these two properties often drive other structural and dynamical properties in empirical citation networks \cite{zeng2017}, we hypothesize that they influence gender imbalance in citations received by papers.
Investigating this hypothesis necessitates reference models that randomize citations while preserving these properties in the given citation network.
However, existing reference models are not intended to investigate associations between citation imbalances and these properties.


\noindent
{\bf Contribution.} The present study makes three significant contributions.
First, we construct a new citation network based on the OpenAlex~\cite{priem2022} and DBLP~\cite{dblp} databases.
Our dataset encompasses rich metadata of papers, such as authors' gender and conference ranks.
Second, we develop a family of reference models for quantifying citation imbalances.
While the existing reference model preserves the number of citations made by each paper \cite{dworkin2020}, our models additionally preserve up to homophily in citations and heterogeneity in the number of citations received per paper.
Third, we quantify gender imbalance in citations between papers published in computer science conferences and examine how network properties influence this gendered citation imbalance.
Our key findings are as follows.
\begin{itemize}
\item Conference papers written by female authors as the first and/or last authors are less likely to receive citations than expected by reference models.
\item Homophily in citations exhibits a strong association with gendered citation imbalance, whereas heterogeneity in the number of citations received per paper shows a minor association.
\item Gendered citation imbalance is most pronounced among conference papers published in the highest-ranked conferences and persists across different subfields.
\item Conference papers written by female authors as the first and/or last authors are less likely to appear at the top of citation-based rankings of papers than expected by reference models.
\end{itemize}

\subsection{Related work}

\subsubsection{Quantifying gender imbalance in academia}

Gender imbalance in academia has been quantitatively investigated using bibliometric data \cite{sugimoto2023}.
Previous studies have focused on gender imbalances within individual disciplines because the extent of gender imbalance and research practices, including authorship, publication venue, and citation, may depend on the discipline.
Dworkin et al.~investigated gender imbalance in citations between journal papers within the neuroscience discipline \cite{dworkin2020}.
Teich et al.~conducted similar analyses within the physics discipline \cite{teich2022}.
Jadidi et al.~examined gender imbalance in research careers, productivity, and collaboration within the computer science discipline \cite{jadidi2018}.
In our study, we investigate gendered citation practices within the computer science discipline, aiming to expand upon previous findings regarding gender imbalance in academia.

Various aspects of gender imbalance in academia have been investigated.
Huang et al.~found that gender imbalance in the productivity of individual researchers is sufficiently explained by that in the length of their publishing careers \cite{huang2020}.
Li et al.~found that gender imbalance in the research performance of researchers with publishing careers of at least 15 years is largely explained by that in the number of collaborators \cite{li2022}.
Gender imbalance in citations is partially explained by gendered patterns in collaborators of authors and the proximity of research subfields between papers \cite{dworkin2020, teich2022}.
In contrast to these studies, we investigate the extent to which homophily in citations and heterogeneity in the number of citations received per paper explain gendered citation imbalance in computer science. 

\subsubsection{Applications of reference models for citation networks}

In general, reference models generate instances that preserve certain properties of the original network and randomize other properties \cite{orsini2015, nakajima2022}. 
Standard practice in network analysis involves comparing the structure and dynamics of the original network with those of randomized instances generated by reference models \cite{orsini2015, nakajima2022}.
Previous studies have deployed reference models to characterize the structure of citation networks. 
For example, Uzzi et al.~investigated atypical combinations of citations associated with the impact of a paper by randomizing citations via a Monte Carlo algorithm \cite{uzzi2013}.
Kojaku et al.~examined anomalous citation patterns among journals using a reference model that accounts for scientific communities and journal size \cite{kojaku2021}.
These reference models preserve the number of citations received by each paper and, thus, are not intended to quantify imbalances in citations received by papers.
Dworkin et al.~introduced a reference model for quantifying gender imbalance in citations received by papers.
This reference model, however, does not preserve properties other than the number of citations made by each paper.
In light of these limitations, we develop a family of reference models that preserve up to the number of citations made by each paper, homophily in citations, and heterogeneity in the number of citations received per paper. 

\section{New dataset}

We construct a citation network composed of conference papers within the computer science discipline.
To this end, we use two open databases of publication records: OpenAlex and DBLP. 
OpenAlex provides extensive coverage of publications and citations between them, while DBLP offers precise conference metadata for computer science publications.
By integrating data from both sources, we construct a publication dataset with enriched metadata.

\subsection{Existing data sources}

\subsubsection{OpenAlex}
OpenAlex provides hundreds of millions of publication records with rich metadata across various disciplines~\cite{priem2022}.
We used publication data from a snapshot of OpenAlex released on February 27th, 2024, extracting 103,335,085 research articles published between 2000 and 2020. 
For each paper $v$, the title, publication date, primary research topic, authors' information (i.e., OpenAlex ID, name, and affiliations), papers cited by $v$, and author position (i.e., first, middle, or last) are available.
Each paper is assigned one of the 4,516 research topics as the primary research topic based on its metadata (e.g., title, abstract, and citations). 
Each research topic belongs to one of the 252 research subfields \cite{openalex_topic_classification}.
OpenAlex uses a proprietary algorithm to assign a unique ID to each author~\cite{priem2022}, and the algorithm identified 45,191,646 authors.
While OpenAlex covers a large extent of publication metadata \cite{lorena2024}, we observed that conference names in OpenAlex data are often missing or not disambiguated for conference papers in computer science.

\subsubsection{DBLP}
DBLP provides millions of publication records in computer science~\cite{dblp}.
We used publication data from a snapshot of DBLP released on April 1st, 2024, and extracted 2,583,968 conference papers published between 2000 and 2020. 
Each paper has the title, publication year, authors' names, and conference names. 
While DBLP covers a large extent of conference names and curates them \cite{dblp_curation}, we observed that it offers less metadata for publications compared to OpenAlex. 
Specifically, DBLP does not provide research subfields and topics for publications.

\subsection{Construction of a new dataset}

We integrate publication data from OpenAlex and DBLP to complement their respective limitations.
We constructed a citation network composed of 99,329 papers (i.e., nodes) and 152,598 citations (i.e., directed edges) between them. 
Table~\ref{table:dataset} shows basic statistics of the dataset.
See Appendix A for additional statistics.
Below, we describe the process of constructing our dataset.

\subsubsection{Country of affiliation and gender of authors}

We assign the country of affiliation and gender to authors in the OpenAlex data using the methods outlined in \cite{nakajima2023}.
First, we assigned a country of affiliation to each author based on the most frequently appearing country in the affiliations of their publication records.
Next, we assigned gender to each author based on their country of affiliation, first publication year, and first name.
We used the Gender API\footnote{\url{https://gender-api.com/en/} (Accessed April 2024)} to infer the gender of authors. 
We set the same parameters as \cite{nakajima2023} for assigning gender based on the outputs of the API.
We then categorized each paper into one of four gender categories (i.e., `MM', `MW', `WM', and `WW') based on the gender of the first and last authors, where the first letter, M or W, indicates that the first author is a man or woman, respectively, and the second letter indicates the gender of the last author.
Note that we categorized sole-author papers by men and women as MM and WW papers, respectively.

\subsubsection{Conference rank}

We used the CORE ranking data in 2021\footnote{\url{https://portal.core.edu.au/conf-ranks/} (Accessed April 2024)} to identify the rank of conferences.
This conference ranking is based on citation statistics of papers and authors presented at the conference and the acceptance rate for the conference.
The conferences are classified into four ranks: `$\text{A}^*$,' `A,' `B,' and `C', with $\text{A}^*$ being the highest and C the lowest. The ranking includes 768 conference names in our dataset.
We manually assigned one of these four ranks to each of the conference names appearing in the DBLP data.

\subsubsection{Matching publications between OpenAlex and DBLP}
We complete the conference metadata for publications in the OpenAlex data by matching publications between OpenAlex and DBLP based on the following criteria: 
(i) both papers are published in the same year, 
(ii) both papers have identical sets of author last names\footnote{We regarded the last space-separated word of a given author's name as their last name \cite{nakajima2023}.}, 
and (iii) the titles of both papers differ by no more than 25\% in terms of the Levenshtein distance between the two titles divided by the length of the longer title~\cite{huang2020}.
If the metadata between OpenAlex and DBLP is inconsistent, we use OpenAlex metadata except for the conference name.


\subsubsection{Citations}

\begin{table}
 \begin{center}
   \caption{Summary of our dataset.}
   \label{table:dataset}
\begin{tabular}{|l|c|} \hline
   \multicolumn{1}{|c|}{Meaning} & Count \\\hline
   Papers & 99,329 \\
   Citations & 152,598 \\ 
   Gender-assigned authors & 83,231 \\ 
   Conferences & 637 \\ 
   Countries of affiliation & 102 \\ 
   Research topics & 1,351 \\ 
   Research subfields & 187 \\ \hline
   $\text{A}^*$-ranked conferences & 59 \\ 
   A-ranked conferences & 124 \\ 
   B-ranked conferences & 257 \\ 
   C-ranked conferences & 197 \\ \hline
   Female authors & 13,097 \\ 
   Male authors & 70,134 \\ \hline
   MM papers & 76,562 \\ 
   MW papers & 8,281 \\ 
   WM papers & 10,479 \\ 
   WW papers & 4,007 \\
   \hline
 \end{tabular}
 \end{center}
\end{table}

Given an observed citation from paper $u$ to paper $v$, we say that $u$ made the citation to $v$, and $v$ received the citation from $u$.
We removed citations made by $u$ to $v$ if either of the following holds true: (i) the publication date of $v$ is ten years older than that of $u$, (ii) both first and last authors of $v$ are either the first or last author of $u$.
The first criterion addresses the different dynamics of citations received by different papers \cite{sinatra2016}; the second criterion avoids self-citations \cite{nakajima2023}.
We also removed papers that neither make nor receive any citations.

\section{Reference models for citation networks}

We aim to characterize gender imbalance in citations received by conference papers in computer science. 
To this end, we compare gendered citation patterns in the original network with those in reference models. 
Our reference models randomize citations made by each paper while preserving certain structural properties of the original network. 

Below, we first describe an existing reference model that preserves the number of citations made by each paper \cite{dworkin2020}.
Then, we extend this model to reference models that preserve up to the number of citations made by each paper, homophily in citations, and heterogeneity in the number of citations received per paper.
We list the properties that each reference model preserves in Table~\ref{table:2}.

\begin{table*}
 \begin{center}
   \caption{Properties to be preserved in each reference model. }
   \label{table:2}
\begin{tabular}{|l|A|} \hline
   \multicolumn{1}{|c|}{Model} & \multicolumn{1}{c|}{Properties to be preserved} \\ \hline
   Random-draws model & $\bullet$ Number of citations made by each paper \\ \hline
   \multirow{2}{*}{Homophilic-draws model} & $\bullet$ Number of citations made by each paper \\
    & $\bullet$ Homophilic citation patterns in terms of given properties of the paper\\ \hline
   \multirow{3}{*}{Preferential-draws model} & $\bullet$ Number of citations made by each paper \\
   & $\bullet$ Homophilic citation patterns in terms of given properties of the paper\\
   & $\bullet$ Heterogeneity in the number of citations received per paper\\
   \hline
 \end{tabular}
 \end{center}
\end{table*}

\subsection{Notation}

We denote by $V = \{v_1, \ldots, v_N\}$ the set of all the papers (i.e., nodes), $E$ the set of citations (i.e., directed edges) in the network, and $M=|E|$ the number of citations, where $N$ is the number of papers.
We denote by $(A_{ij})_{1 \leq i \leq N, 1 \leq j \leq N}$ an adjacency matrix of the network, where $A_{ij} = 1$ if $(v_i, v_j)$ is in $E$ and $A_{ij} = 0$ otherwise.
We denote by $k_{i} = \sum_{j=1}^N A_{ij}$ the number of citations made by $v_i \in V$, and we denote by $c_{j} = \sum_{i=1}^N A_{ij}$ the number of citations received by $v_j \in V$.

\subsection{Random-draws model}

We first describe the random-draws model \cite{dworkin2020}.
Consider an observed citation made by paper $v_i \in V$ to paper $v_{i'} \in V$. 
This model assumes that $v_{i'}$ is drawn uniformly at random from a set of papers that $v_i$ could potentially cite, which we denote by $\overline{V}_{\text{RD}}(i)$.
The subscript `RD' indicates the `random-draws' model.
According to our definition of citations between papers, we define $\overline{V}_{\text{RD}}(i)$ as the set of papers that meet the following criteria: 
(i) the publication date of each paper is at most ten years older than that of $v_i$, and (ii) both the first and last authors of each paper are neither the first nor last authors of $v_i$.
Then, any paper in $\overline{V}_{\text{RD}}(i)$ receives a citation from $v_{i}$ with the probability $1 / |\overline{V}_{\text{RD}}(i)|$ under the random-draws model.
Therefore, the probability of a citation made by $v_i$ to $v_j \in V$ under the model is given by
\begin{align}
\bar{w}_{ij, \text{RD}} = 
\begin{cases}
k_i / |\overline{V}_{\text{RD}}(i)| & \text{if }v_j \in \overline{V}_{\text{RD}}(i), \\
0 & \text{otherwise}.
\end{cases}
\label{eq:1}
\end{align}
Then, aggregating the probability of a citation received by $v_j$ from other papers yields the expected number of citations received by $v_j$ under the model:
\begin{align}
\bar{c}_{j, \text{RD}} = \sum_{i=1}^N \bar{w}_{ij, \text{RD}}.
\label{eq:2}
\end{align}
The random-draws model preserves the expected number of citations made by each paper in the original network.
Indeed, it holds true that $\sum_{j=1}^N \bar{w}_{ij, \text{RD}} = k_i$ for any $v_i \in V$.
See Algorithm \ref{alg:1} for the pseudocode of the random-draws model.

\subsection{Homophilic-draws model}
Empirical citation networks often exhibit homophily, where papers tend to cite those with similar properties \cite{ciotti2016}. 
Homophily in citations may be associated with gender imbalance in citations received by papers \cite{nakajima2023}.
To investigate this possibility, we need to compare gendered citation patterns between a reference model that destroys homophilic citation patterns in the original network and one that preserves them.
The random-draws model destroys homophilic citation patterns.
Therefore, we extend the random-draws model to a reference model that preserves homophilic citation patterns in the original network, which we refer to as the homophilic-draws model.

We denote by $S$ a set of properties of the paper, other than the gender category, that can be relevant to homophilic citations.
In this study, we define the set of the following three properties as $S$: (i) conference rank, (ii) country of affiliation, and (iii) research topic.
Note that $S$ is not limited to these properties.
For example, one may focus on journal names for citation networks in other disciplines \cite{kojaku2021}.

We describe the algorithmic procedure.
Consider an observed citation made by $v_i \in V$ to $v_{i'} \in V$. 
We define $\overline{V}_{\text{HD}}(i, i', S)$ as the set of papers in $\overline{V}_{\text{RD}}(i)$ that belong to the same category\footnote{For example, in our analyses, $\overline{V}_{\text{HD}}(i, i', S)$ is defined as the set of papers in $\overline{V}_{\text{RD}}(i)$ that belong to the same conference rank, the same country of affiliation, and the same research topic as $v_{i'}$.} as $v_{i'}$ for any property in $S$.
Note that $\overline{V}_{\text{HD}}(i, i', S)$ includes $v_{i'}$.
The subscript `HD' indicates the `homophilic-draws' model.
We assume that any paper in $\overline{V}_{\text{HD}}(i, i', S)$ receives a citation from $v_{i}$ with the probability $1 / |\overline{V}_{\text{HD}}(i, i', S)|$ under the homophilic-draws model.
Then, the probability of a citation made by $v_i$ to $v_j \in V$ under the model is given by
\begin{align}
\bar{w}_{ij, \text{HD}} = \sum_{\substack{i'=1 \\ (v_{i}, v_{i'}) \in E \\ v_j \in \overline{V}_{\text{HD}}(i, i', S)}}^N 1 / |\overline{V}_{\text{HD}}(i, i', S)|.
\label{eq:3}
\end{align}
The expected number of citations received by $v_{j} \in V$ under the model is given by
\begin{align}
\bar{c}_{j, \text{HD}} = \sum_{i=1}^N \bar{w}_{ij, \text{HD}}.
\label{eq:4}
\end{align}
The homophilic-draws model preserves the number of citations made by each paper. 
In addition, it preserves the number of citations between papers in terms of any pair of conference ranks, any pair of countries of affiliation, and any pair of research topics (see Figs.~\ref{fig:1}(b)--\ref{fig:1}(d) for numerical evidence).
See Algorithm \ref{alg:2} for the pseudocode of the homophilic-draws model.

\subsection{Preferential-draws model}
Empirical citation networks may exhibit a greater extent of heterogeneity in the number of citations received per paper than expected under the homophilic-draws model.
One possible explanation is that citation dynamics largely adhere to the preferential attachment mechanism (i.e., papers are more likely to cite papers that receive more citations in a research topic) \cite{eom2011}.
We investigate the extent to which heterogeneity in the number of citations received per paper is associated with gender imbalance in citations received by papers.
To this end, we extend the homophilic-draws model to a reference model that approximately preserves heterogeneity in the number of citations received per paper, which we refer to as the preferential-draws model. 

We examine citations made by older papers in sequential order.
We sort the papers by publication date in ascending order: $v_{x_1}, \ldots, v_{x_N}$, where $x_l$ is the index of the $l$-th paper in the sorted result; $v_{x_1}$ and $v_{x_N}$ are the oldest and newest papers, respectively.
We randomize citations made by paper $v_{x_l}$ by accounting for the following properties for the papers $v_{x_1}, \ldots, v_{x_{l-1}}$: (i) any property in $S$ and (ii) the expected number of citations received before $v_{x_l}$ is published under the model.
Below, we denote by $\bar{c}_{i, l, \text{PD}}$ the expected number of citations received by $v_i \in V$ from the papers $v_{x_1}, \ldots, v_{x_l}$ under the model, where the subscript `PD' indicates the `preferential-draws' model.

We describe the algorithmic procedure.
First, we initialize $\bar{c}_{i, l, \text{PD}}$ to zero for any $i= 1, \ldots, N$ and any $l = 1, \ldots, N$.
Second, we perform the following procedure from $l=2$ to $N$ in sequential order.
Consider an observed citation made by paper $v_{x_l}$ to $v_{i'} \in V$. 
We define $\overline{V}_{\text{PD}}(l, i', S)$ as the set of papers in $\overline{V}_{\text{HD}}(x_l, i', S)$ that receives the same expected number of citations from the papers $v_{x_1}, \ldots, v_{x_{l-1}}$ as $v_{i'}$, i.e., $\overline{V}_{\text{PD}}(l, i', S) = \{v_j\ |\ v_j \in \overline{V}_{\text{HD}}(x_l, i', S) \ \land \ \bar{c}_{j, l-1, \text{PD}} = \bar{c}_{i', l-1, \text{PD}}\}$.
Note that $\overline{V}_{\text{PD}}(l, i', S)$ includes $v_{i'}$.
We assume that any paper in $\overline{V}_{\text{PD}}(l, i', S)$ receives a citation from $v_{x_{l}}$ with the probability $1 / |\overline{V}_{\text{PD}}(l, i', S)|$.
Therefore, for any paper $v_j \in \overline{V}_{\text{PD}}(l, i', S)$ and any $z = l, \ldots, N$, we increase $\bar{c}_{j, z, \text{PD}}$ by $1 / |\overline{V}_{\text{PD}}(l, i', S)|$.
Finally, the probability of a citation made by $v_i$ to $v_j \in V$ under the model is given by
\begin{align}
\bar{w}_{ij, \text{PD}} = \sum_{\substack{l=2 \\ x_l=i}}^N \sum_{\substack{i'=1 \\ (v_{i}, v_{i'}) \in E \\ v_j \in \overline{V}_{\text{PD}}(l, i', S)}}^N 1 / |\overline{V}_{\text{PD}}(l, i', S)|.
\label{eq:5}
\end{align}
The expected number of citations received by $v_{j} \in V$ under the model is given by
\begin{align}
\bar{c}_{j, \text{PD}} = \sum_{i=1}^N \bar{w}_{ij, \text{PD}}.
\label{eq:6}
\end{align}
As with the homophilic-draws model, the preferential-draws model preserves the number of citations made by each paper and homophilic citation patterns in terms of the conference rank, country of affiliation, and research topic.
In addition, it approximately preserves the heterogeneity in the number of citations received per paper in the original network (see Fig.~\ref{fig:1}(e) for numerical evidence).
See Algorithm \ref{alg:3} for the pseudocode of the preferential-draws model. 

\section{Characterizing gender imbalance in citations}

We quantify gender imbalance in citations received by papers using the family of reference models.
Consider two subsets of papers, denoted by $V_{\text{from}} \subseteq V$ and $V_{\text{to}} \subseteq V$. 
We measure the extent to which the papers in $V_{\text{from}}$ over- or under-cite the papers in $V_{\text{to}}$ and in a given gender category $g \in \{\text{MM}, \text{MW}, \text{WM}, \text{WW}\}$ \cite{dworkin2020, teich2022}.
We first count the citations made by the papers in $V_\text{from}$ to those in $V_{\text{to}}$ and in gender category $g$, which we denote by $n_{g, \text{obs}}$. 
Then, we compare $n_{g, \text{obs}}$ with the expectation obtained by a reference model.
The expectation under the random-draws model is given by  
\begin{align}
\bar{n}_{g, \text{RD}} = \sum_{v_i \in V_{\text{from}}} k_i' p_{i, g, \text{RD}},
\label{eq:7}
\end{align}
where $k_i'$ is the number of citations made by $v_i$ to the papers in $V_{\text{to}}$, and $p_{i, g, \text{RD}}$ is the fraction of the papers that are in both $\overline{V}_{\text{RD}}(i)$ and $g$.
The expectation under the homophilic-draws model is given by
\begin{align}
\bar{n}_{g, \text{HD}} = \sum_{v_i \in V_{\text{from}}} \sum_{\substack{v_j \in V_{\text{to}}\\(v_i, v_j) \in E}} p_{i,j,S,g, \text{HD}},
\label{eq:8}
\end{align}
where $p_{i,j,S,g, \text{HD}}$ is the fraction of the papers that are in both $\overline{V}_{\text{HD}}(i, j, S)$ and $g$.
The expectation under the preferential-draws model is given by
\begin{align}
\bar{n}_{g, \text{PD}} = \sum_{l=2}^N \sum_{\substack{v_j \in V_{\text{to}}\\(v_{x_l}, v_j) \in E}} p_{l,j,S,g,\text{PD}},
\label{eq:9}
\end{align}
where $p_{l,j,S,g,\text{PD}}$ is the fraction of the papers that are in both $\overline{V}_{\text{PD}}(l, j, S)$ and $g$.
For given reference model $r \in \{\text{RD}, \text{HD}, \text{PD}\}$, we calculate the over/under-citation made by the papers in $V_{\text{from}}$ to those in $V_{\text{to}}$ as $(n_{g, \text{obs}} - \bar{n}_{g, r}) / \bar{n}_{g, r}$.

We also perform bootstrap resampling to estimate confidence intervals for the over/under-citations made by papers in our dataset (see Appendix C for details).

\section{Characterizing gender imbalance in citation-based rankings}

We quantify gender imbalance in citation-based rankings of papers.
To this end, we compute the fraction of W\textbar{}W papers (i.e., MW, WM, and WW papers) among the top $d$\% of papers with the largest values of a given impact measure \cite{karimi2018}.
We compare this fraction in the original network with that in the reference models.

We first use the number of citations received by the paper, $c_j$, as the impact measure \cite{zeng2017}.
We employ Eqs.~\eqref{eq:2}, \eqref{eq:4}, and \eqref{eq:6} to compute the ranking of the papers based on the expectations for the reference models.

The quantity $c_j$ does not account for which papers cited $v_j$ and how $v_j$ influenced the impact of those citing papers.
Alternative metrics exist for measuring the impact of a paper \cite{zeng2017}. 
One such metric is PageRank, originally introduced for ranking web pages \cite{brin1998}, and it has been deployed in various applications, including the ranking of papers in several disciplines \cite{ma2008, masuda2017}.
We also use PageRank as the impact measure for the papers in our study.

We compute the PageRank of the papers in the original network as follows.
Consider a random walk with teleportations on the network: an individual reading paper $v_i \in V$ follows one of the citations made by $v_i$ uniformly at random with the probability $\alpha$ and teleports to one of the $N$ research papers with probability $1 - \alpha$.
If the individual reads a paper that makes no citations, they also teleport to one of the $N$ research papers.
We assume that the paper to which the individual teleports is selected from the probability distribution proportional to the number of citations received by the paper \cite{lambiotte2012}.
The PageRank of $v_i$, denoted by $p^*_{i, \alpha}$, is the probability that the individual reads paper $v_i$ after a sufficient number of time steps. 
We define $p^*_{i, \alpha}$ by the following equation \cite{lambiotte2012}
\begin{align}
p^*_{i, \alpha}= &(1-\alpha) \frac{c_{i}}{M} + \alpha \sum_{\substack{j=1}}^N T_{ji} p^*_{j, \alpha},
\label{eq:10}
\end{align}
where we define
\begin{align}
T_{ij}=
\begin{cases}
A_{ij}/k_i & \text{if }k_i > 0, \\
c_j / M & \text{otherwise}. \\
\end{cases}
\label{eq:11}
\end{align}
We refer to $(T_{ij})_{1 \leq i \leq N, 1 \leq j \leq N}$ as a transition-probability matrix for the original network.
We set $\alpha = 0.85$ \cite{brin1998}.
We compute the PageRank of the papers iteratively as follows.
We assume that the initial probability distribution is $p_i(0) = c_{i} / M$ for any $i = 1, \ldots, N$.
The computation at each step $t > 0$ yields
\begin{align}
p_i(t+1) = &(1-\alpha) \frac{c_{i}}{M} + \alpha \sum_{j=1}^N T_{ji} p_j(t).
\label{eq:12}
\end{align}
The computation ends when $t$ reaches $t_{\text{max}}$ or it holds that
\begin{align}
\frac{1}{N} \sum_{i=1}^N |p_i(t+1) - p_i(t)| < \epsilon,
\label{eq:13}
\end{align}
where we set $t_{\text{max}} = 100$ and $\epsilon = 10^{-6}$ \cite{networkx_pagerank}.

To compute the PageRank of the papers for a given reference model, we assume the following: (i) an individual reading paper $v_i \in V$ switches to paper $v_j \in V$ with the probability of a citation made by $v_i$ to $v_j$ under the reference model; (ii) the paper to be teleported is selected from the probability distribution proportional to the expected number of citations received by the paper under the reference model.
Based on these assumptions, we define the PageRank of $v_i$ for the reference model, denoted by $\bar{p}^*_{i, \alpha, r}$, as
\begin{align}
\bar{p}^*_{i, \alpha, r} = &(1-\alpha) \frac{\bar{c}_{i, r}}{M} + \alpha \sum_{j=1}^N \overline{T}_{ji, r}\, \bar{p}^*_{j, \alpha, r},
\label{eq:14}
\end{align}
where $r \in \{\text{RD}, \text{HD}, \text{PD}\}$ and we define
\begin{align}
\overline{T}_{ij, r}=
\begin{cases}
\bar{w}_{ij, r} / k_{i} & \text{if }k_i > 0, \\
\bar{c}_{j, r} / M & \text{otherwise}. \\
\end{cases}
\label{eq:15}
\end{align}
For any $r \in \{\text{RD}, \text{HD}, \text{PD}\}$, it holds true that $\sum_{j=1}^N \bar{w}_{ij, r} = k_i$ for any $v_i \in V$ and $\sum_{j=1}^N \bar{c}_{j, r} = M$.
We refer to $(\overline{T}_{ij, r})_{1 \leq i \leq N, 1 \leq j \leq N}$ as a transition-probability matrix for the reference model.
Note that $(\overline{T}_{ij, r})_{1 \leq i \leq N, 1 \leq j \leq N}$ for any $r \in \{\text{RD}, \text{HD}, \text{PD}\}$ is typically much denser than $(T_{ij})_{1 \leq i \leq N, 1 \leq j \leq N}$.
We assume that the initial probability distribution is $\bar{p}_{i, r}(0) = \bar{c}_{i, r} / M$ for any $v_i \in V$ and any $r \in \{\text{RD}, \text{HD}, \text{PD}\}$. 
We set $\alpha = 0.85$.
The iterative computation procedure is the same as that for the original network.

For both impact measures, we use the score (i.e., the number of citations received by the paper or PageRank) of each paper $v_i$ divided by the average score of the papers that were published in the same year and belong to the same subfield as $v_i$.
This normalization addresses the time-and-subfield dependence of the average number of citations received per paper \cite{radicchi2008}. 

\section{Results} \label{section:3}

\subsection{Comparison of reference models}

\begin{figure*}[t]
\centering
\includegraphics[scale=0.4]{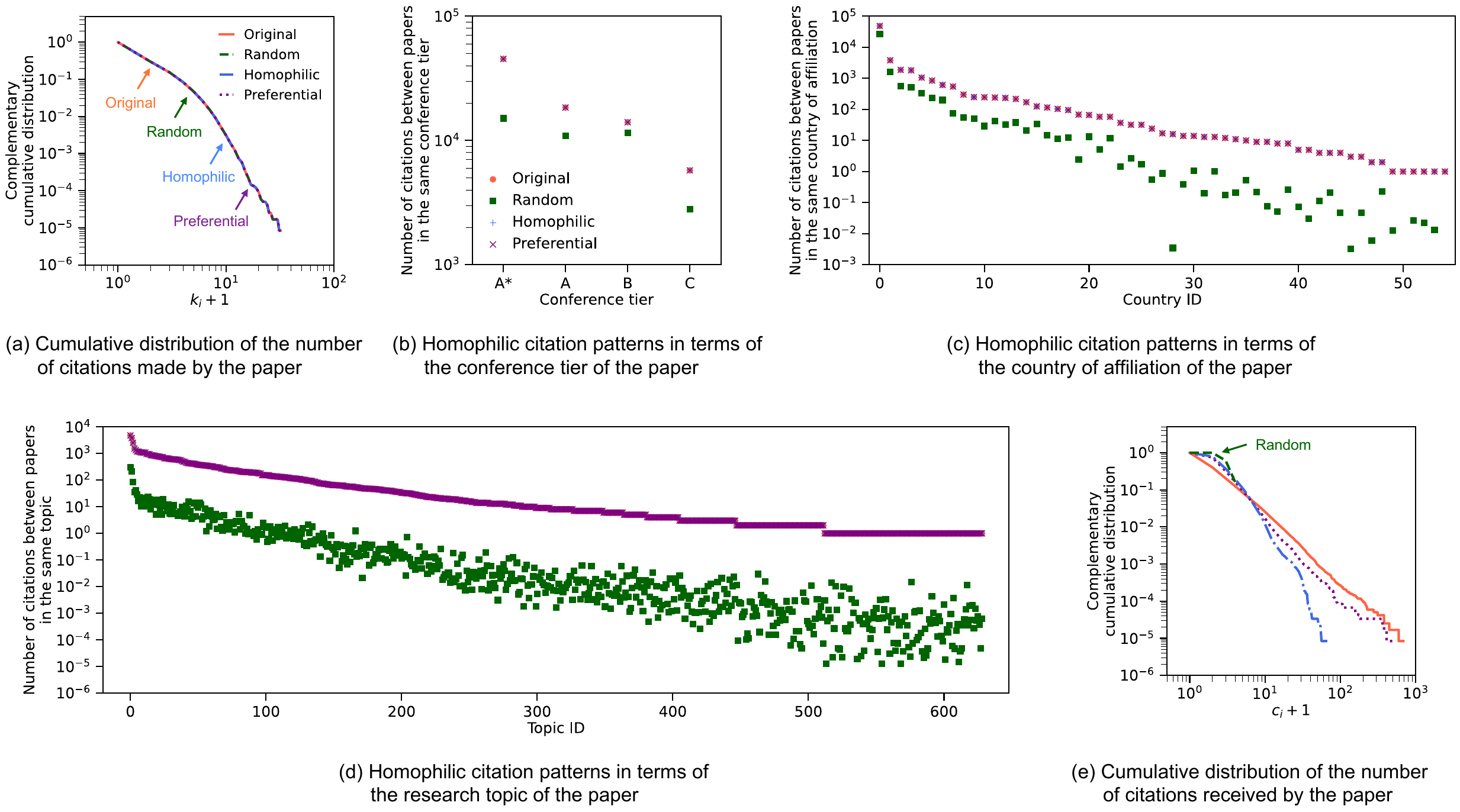}
\caption{
Comparison of structural properties between the original citation network and reference models. The legend `Original' indicates the result for the original network, `Random' indicates that for the random-draws model, `Homophilic' indicates that for the homophilic-draws model, and `Preferential' indicates that for the preferential-draws model. In Figs.~\ref{fig:1}(c) and \ref{fig:1}(d), we focused on the 55 countries of affiliation and the 625 research topics, each of which has one or more homophilic citations in the original network. We sorted the IDs in descending order by the number of homophilic citations in each case. Curves that completely or heavily overlap are indicated by arrows and labels.
}
\label{fig:1}
\end{figure*}

We first examine which structural properties in the original citation network are preserved by the three reference models.
First, all models exactly preserve the distribution of the number of citations made by the paper in the original network (see Fig.~\ref{fig:1}(a)).
Second, while the random-draws model destroys homophilic citation patterns in the original network, both the homophilic-draws and preferential-draws models exactly preserve (see Figs.~\ref{fig:1}(b)--(d)).
Third, while the random-draws and homophilic-draws models hardly preserve the heterogeneity in the number of citations received per paper in the original network, the preferential-draws model approximately preserves (see Fig.~\ref{fig:1}(e)).
These results align with our expectations.

\subsection{Gendered citation imbalance in conference papers}

\begin{figure*}[t]
\centering
\includegraphics[scale=0.3]{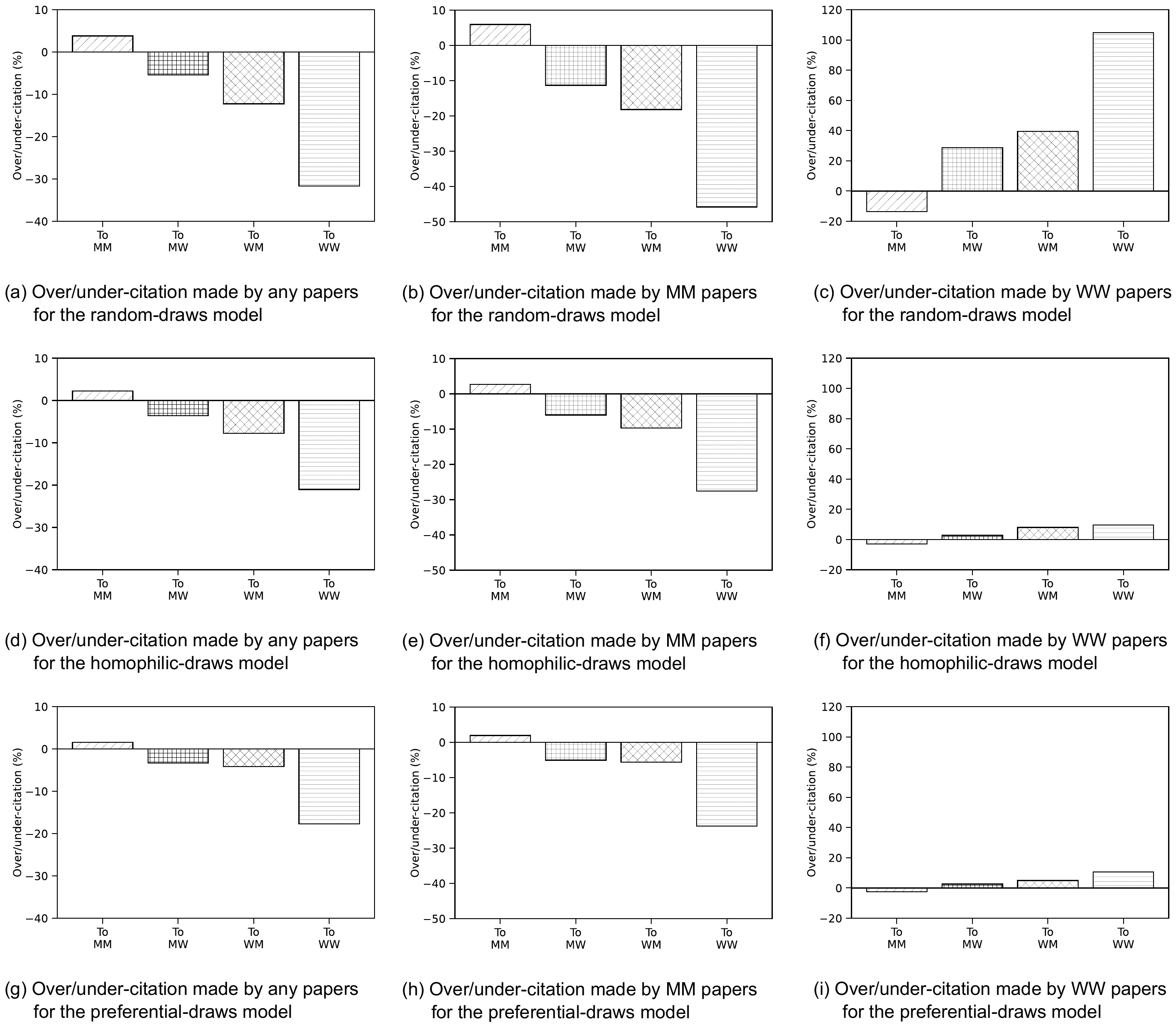}
\caption{
Gender imbalance in citations received by conference papers in computer science.
}
\label{fig:2}
\end{figure*}

We now quantify gender imbalance in citations between papers published in computer science conferences.
Each conference paper is categorized into one of four gender categories (i.e., MM, MW, WM, or WW) based on the gender of the first and last authors.
We first count the number of citations made by a subset of papers to those in each gender category. 
Then, we compare the obtained citation counts with the expectations derived from the reference models.

Figure \ref{fig:2}(a) shows the over/under-citation made by any papers to the papers in each gender category, computed using the random-draws model.
We found that any papers tend to over-cite MM papers and under-cite MW, WM, and WW papers. 
Using the random-draws model, we also computed the over/under-citation made by MM and WW papers.
The over/under-citation patterns made by MM papers are qualitatively similar to those made by any papers (see Fig.~\ref{fig:2}(b)).
This result is expected because the fraction of citations made by MM papers is dominant (i.e., 77.7\%).
In contrast, WW papers exhibit the opposite over/under-citation patterns: they tend to under-cite MM papers and over-cite MW, WM, and WW papers (see Fig.~\ref{fig:2}(c)).
These results are largely consistent with previous findings on gendered citation imbalances in journal papers in neuroscience and physics \cite{dworkin2020, teich2022}.

Comparisons of the over/under-citation between the random-draws and homophilic-draws models will reveal the influence of homophilic citation patterns.
This is because the random-draws model destroys homophilic citation patterns in the original network, whereas the homophilic-draws model preserves them.
Similarly, comparisons of the over/under-citation between the homophilic-draws and the preferential-draws models will reveal the influence of the heterogeneity in the number of citations received per paper. 

Figures \ref{fig:2}(d)--\ref{fig:2}(f) show the results for the homophilic-draws model.
We found that homophilic citation patterns are associated with the under-citation received by MW, WM, and WW papers.
Specifically, the over/under-citation received by MW papers is reduced from $-4.8\%$ for the random-draws model (see Fig.~\ref{fig:2}(a)) to $-3.3\%$ for the homophilic-draws model (see Fig.~\ref{fig:2}(d)).
Similarly, the over/under-citations received by WM and WW papers are reduced from $-12.9\%$ and $-31.4\%$ for the random-draws model (see Fig.~\ref{fig:2}(a)) to $-8.2\%$ and $-21.2\%$ for the homophilic-draws model (see Fig.~\ref{fig:2}(d)), respectively.
We make qualitatively similar observations for the over/under-citation made by MM papers (see Figs.~\ref{fig:2}(b) and \ref{fig:2}(e)).
Furthermore, we found that gendered citation patterns made by WW papers are sufficiently explained by homophilic citation patterns.
Indeed, the over/under-citation made by WW papers is almost eliminated in the homophilic-draws model (see Figs.~\ref{fig:2}(c) and \ref{fig:2}(f)).

Figures \ref{fig:2}(g)--\ref{fig:2}(i) show the results for the preferential-draws model.
We found that the heterogeneity in the number of citations received per paper is slightly associated with the under-citation received by MW, WM, and WW papers.
We also make qualitatively similar observations for MM papers (see Fig.~\ref{fig:2}(h)).
In contrast, the heterogeneity in the number of citations received per paper contributes little to describing gendered citation patterns made by WW papers (see Figs.~\ref{fig:2}(f) and \ref{fig:2}(i)).

In summary, gender imbalance in citations received by conference papers in computer science exists.
Homophily in citations is strongly associated with the gender imbalance in citations, whereas heterogeneity in the number of citations received per paper shows a minor association with it.
Gendered citation patterns made by MM papers persist even after controlling for both of these properties, whereas those made by WW papers are sufficiently explained by homophilic citation patterns.

\subsection{Gendered citation imbalance across conference ranks and subfields}

Gender imbalance in academia has been investigated in terms of prestige, including institutional prestige \cite{huang2020}.
Journals and conferences are often ranked based on citation statistics of papers published there \cite{garfield1972, freyne2010}.
Indeed, several ranking systems exist for conferences in computer science, with the highest-ranked conferences often regarded as prestigious \cite{li2018}.
We hypothesize that the gendered citation imbalance in computer science is associated with the prestige of the conference.
To investigate this possibility, we computed the over/under-citation received by MM and WW papers within each conference rank using the preferential-draws model. 
It quantifies the over/under-citation that cannot be immediately explained by the number of citations made by each paper, homophily in citations, and the heterogeneity in the number of citations received per paper.

\begin{figure}[t]
\centering
\includegraphics[scale=0.15]{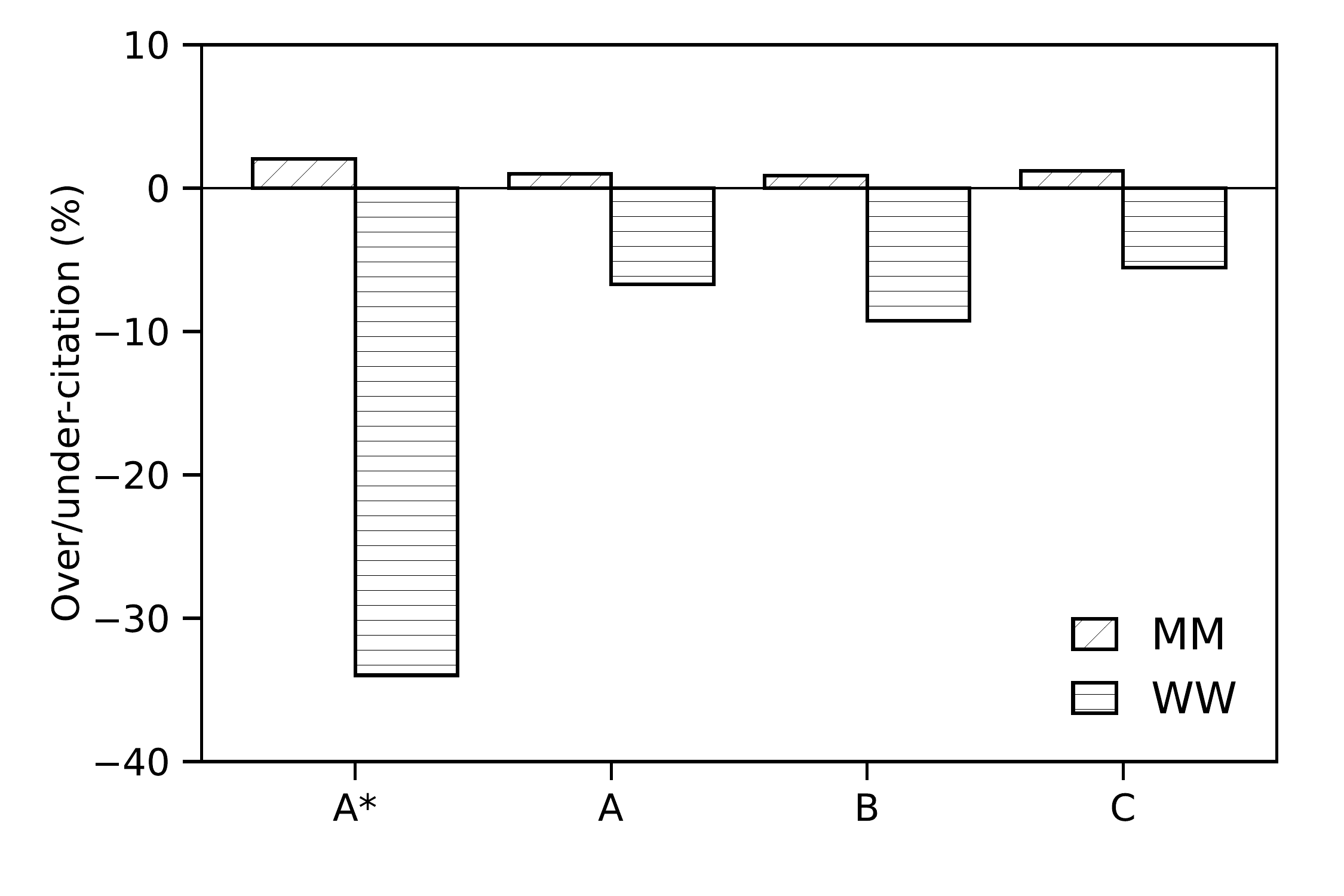}
\caption{
Gendered citation imbalance across conference ranks.
}
\label{fig:3}
\end{figure}

Figure \ref{fig:3} compares the over/under-citation received by MM and WW papers within each conference rank. 
The gender imbalance persists across conference ranks. Notably, the degree of under-citation received by WW papers published in $\text{A}^*$-ranked conferences is at least four times larger than in A or lower-ranked conferences. In contrast, the degree of over-citation received by MM papers does not vary greatly across different conference ranks. The difference in the over/under-citation received by WW papers by conference rank is not immediately explained by the fraction of WW papers within each rank: 2.9\% for $\text{A}^*$ rank, 4.5\% for A rank, 3.7\% for B rank, and 5.8\% for C rank.

The degree of gender imbalance often varies by subfield within a single discipline. 
Indeed, gendered citation imbalance exists across different subfields in physics \cite{teich2022}. 
We compare the degree of gender imbalance in citations across different subfields in computer science. 
To this end, we focus on the 11 subfields categorized under the `Computer Science' discipline in the OpenAlex data. We then compute the degree of over/under-citation received by MM and WW papers within each subfield using the preferential-draws model.

\begin{figure}[t]
\centering
\includegraphics[scale=0.12]{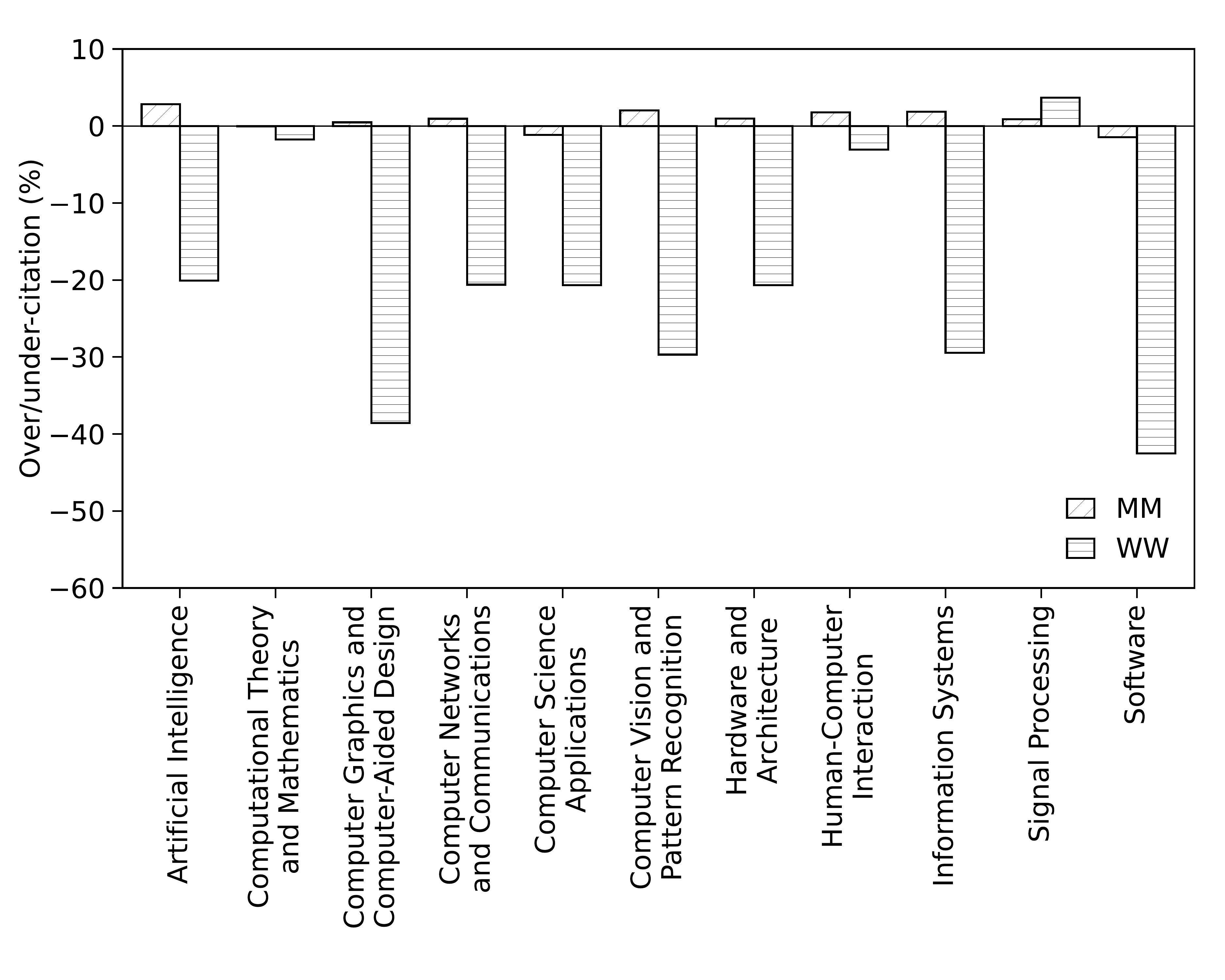}
\caption{
Gendered citation imbalance across subfields.
}
\label{fig:4}
\end{figure}

We found that the under-citation received by WW papers exists in all subfields except for the `signal processing' subfield (see Fig.~\ref{fig:4}).
The variation in the degree of the under-citation received by WW papers across subfields cannot be directly explained by that in the fraction of WW papers in each subfield: the Spearman's rank correlation coefficient is 0.14 ($P$-value is 0.69).

To sum up, gender imbalance in citations received by conference papers persists across conference ranks and subfields in computer science. 
This imbalance is most pronounced for papers published in the highest-ranked conferences.

\subsection{Gender imbalance in citation-based rankings}

\begin{figure*}[t]
\centering
\includegraphics[scale=0.5]{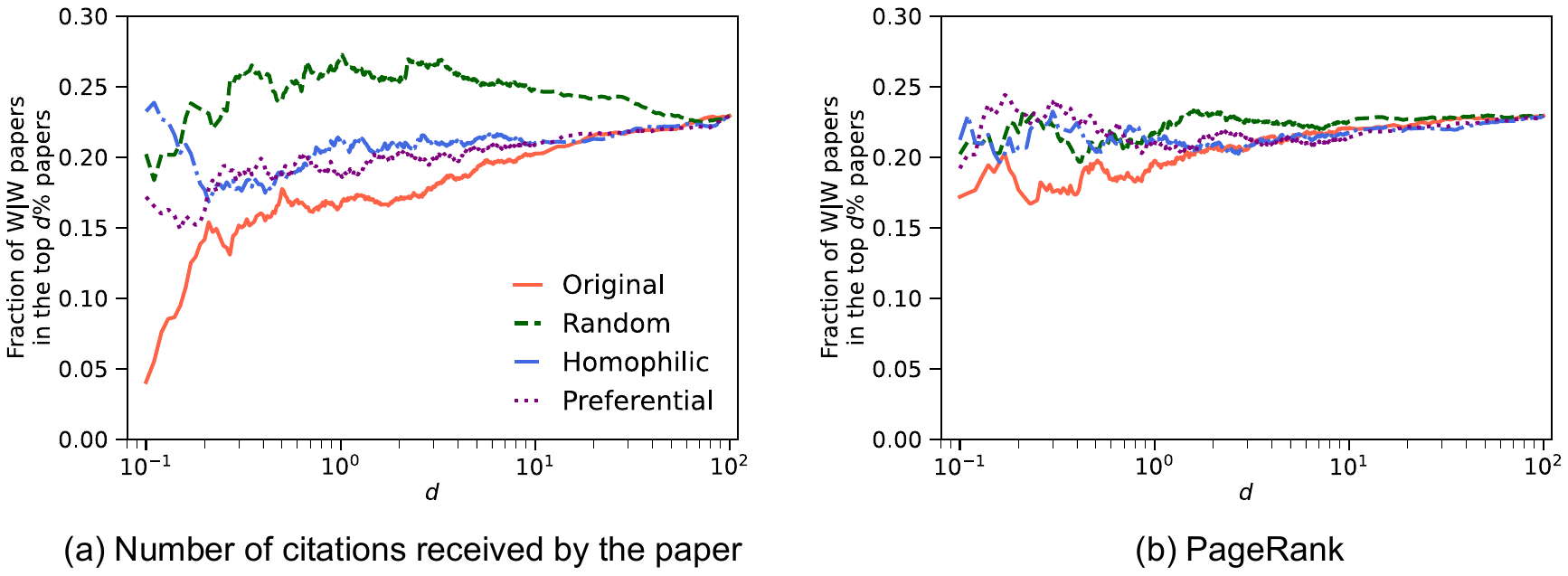}
\caption{
Gender imbalance in citation-based rankings of conference papers.
}
\label{fig:5}
\end{figure*}

We have observed gender imbalance in citations received by conference papers in computer science. 
We further examine how this gendered citation imbalance influences gender imbalance in citation-based rankings of papers. 
To this end, we first rank the conference papers based on the number of citations received. 
We compute the number of citations received by each paper in the original network and compare it with the expectations in a given reference model (see Methods Section for details). 
For a given $d$ value, we then compute the fraction of W\textbar\text{W} papers (i.e., MW, WM, and WW papers) among the top $d$\% of papers with the highest number of citations received, comparing the original network with the reference model.

Figure \ref{fig:5}(a) compares the fraction of W\textbar\text{W} papers in the top $d$\% papers between the original network and the reference models.
We found that the fraction of W\textbar\text{W} papers in the top $d\%$ for the original network is lower than that for any reference model across most $d$ values in the range $[0.1, 20]$.
Additionally, the fraction of W\textbar\text{W} papers for the homophilic-draws model is smaller than that for the random-draws model across most $d$ values in the range $[0.1, 100]$.
This comparison reveals that homophily in citations is associated with gender imbalance in the ranking of papers, consistent with previous findings in the physics discipline \cite{karimi2018}.
In contrast, the fraction of W\textbar\text{W} papers for the preferential-draws model is comparable to that for the homophilic-draws model, indicating that heterogeneity in the number of citations received per paper has a minor influence on gender imbalance in the rankings of papers.

The number of citations received by a given paper $v$ does not account for which papers cited $v$ or how $v$ influenced the impact of the citing papers.
To address this, we use PageRank, which measures the impact of a paper based on the citation network \cite{zeng2017}, to rank the conference papers.
We compute the PageRank of the papers in both the original network and a given reference model (see Methods Section for details).
Then, we compare the fraction of W\textbar\text{W} papers in the top $d$\% of papers with the largest PageRank between the original network and the reference model.

Figure \ref{fig:5}(b) compares the fraction of W\textbar\text{W} papers in the top $d$\% papers between the original network and the reference models.
The fraction for the original network is lower than that for both reference models across most $d$ values in the range $[0.1, 2]$.
Comparisons among the reference models indicate that homophily in citations moderately influences gender imbalance in the ranking of papers based on PageRank, whereas heterogeneity in the number of citations received per paper does little.

To sum up, gender imbalance in citation-based rankings of conference papers in computer science exists.
Homophily in citations is associated with this gender imbalance, whereas heterogeneity in the number of citations received per paper has little association.

\section{Discussion}

\subsection{Gender imbalance in citations received by papers}
We found that conference papers written by female authors as the first and/or last authors are less likely to receive citations across different subfields in computer science.
This result is largely consistent with previous findings on gendered citation imbalance in other disciplines \cite{dworkin2020, fulvio2021, xwang2021, teich2022}.
Additionally, we make significant insights into these previous findings.
First, we found that homophily in citations is strongly associated with gender imbalance in citations, whereas the heterogeneity in the number of citations received per paper shows a minor association.
Second, we found that the prestige of the highest-ranked conferences is strongly associated with the gender imbalance in citations.
Our results indicate the presence of gendered citation patterns in computer science that cannot be explained by these structural properties alone.

Gender homophily has been observed in citations, and it is often associated with gender imbalance in citations \cite{tekles2022, zhou2024}.
Tekles et al.~found that gendered citation patterns in the biological and medical disciplines are sufficiently explained by homophilic citations in terms of the research topic \cite{tekles2022}.
Consistent with this finding, we found that gendered citation patterns for papers written by female authors as the first and last authors in computer science are also largely explained by homophilic citations. 
Therefore, our results support the previous finding that extensively controlling for properties of the paper is important for assessing the degree of gender imbalance in citations \cite{tekles2022}.

\subsection{Gender imbalance in citation-based rankings}
Karimi et al.~found that papers categorized within the "minority" group of the research topic are less likely to appear at the top of citation-based rankings than expected by random chance \cite{karimi2018}.
In contrast, we proposed the use of reference models to assess imbalances in citation-based rankings.
Our findings indicate that papers written by female authors as the first and/or last authors are less likely to appear at the top of citation-based rankings, even after accounting for homophily in citations and heterogeneity in citation counts per paper.

We expect that modifying network structures using reference models could potentially ameliorate gender imbalances in citation-based rankings. 
For example, one could utilize the preferential-draws model to calculate the PageRank of each paper in the given citation network. 
Indeed, our results suggest that adjusting a transition-probability matrix and a probability distribution of teleportation in the PageRank using the preferential-draws model improves gender imbalance in the ranking of papers (see Fig.~\ref{fig:5}(b)). 
A similar approach has been employed to enhance the fairness of PageRank among nodes in a network \cite{tsioutsiouliklis2021}. 
In contrast to this previous study \cite{tsioutsiouliklis2021}, our reference models do not require any algorithmic parameters to enhance the fairness in node rankings.

The development of fairness-aware methods for network analysis has emerged as a recent and challenging topic \cite{saxena2024}.
The prospect of ameliorating imbalances in node rankings by modifying network structures using reference models remains an open area for exploration and discussion. 
This approach may be extended to encompass other centrality measures and network types and is expected to contribute to the establishment of fairness-aware methods in node rankings.

\section{Conclusion}

In this study, we quantified gendered citation imbalance in conference papers in computer science.
Previous studies have identified gender imbalance in the number of researchers \cite{way2016, jadidi2018, huang2020, laberge2022}, research career \cite{jadidi2018, huang2020, morgan2021, lietz2024}, and authorship \cite{holman2018, llwang2021} within the computer science discipline.
Additionally, studies have quantified gendered citation imbalance in journal papers across other disciplines \cite{dworkin2020, fulvio2021, xwang2021, teich2022}. 
By developing a family of reference models for citation networks, we extended previous findings on gender imbalance in computer science in terms of citation practices.

Our study presents several future challenges.
Our reference models do not consider the attractiveness or fitness of the paper \cite{eom2011, wang2013} and the authors' reputation \cite{petersen2014}, both of which significantly impact citation dynamics.
Extending our reference models to include these factors could help us understand how these properties are associated with gendered citation imbalance. 
Furthermore, citation imbalances may also be present concerning other author attributes, such as nationality \cite{nakajima2023} and race \cite{kozlowski2022, liu2023}, in addition to gender, within the computer science discipline. Investigating citation imbalances in relation to these other relevant author attributes warrants further work.

\section{Acknowledgments}

This work was supported by Japan Science and Technology Agency (JST) as part of Adopting Sustainable Partnerships for Innovative Research Ecosystem (ASPIRE), Grant Number JPMJAP2328. 
KN thanks the financial support by JSPS KAKENHI Grant Number 24K21056.
YS thanks the support by JST Presto Grant Number JPMJPR21C5.

\bibliography{main}

\newpage
\appendix
\section{Appendix A: Additional statistics for our dataset}

\begin{figure}[h]
\centering
\includegraphics[scale=0.25]{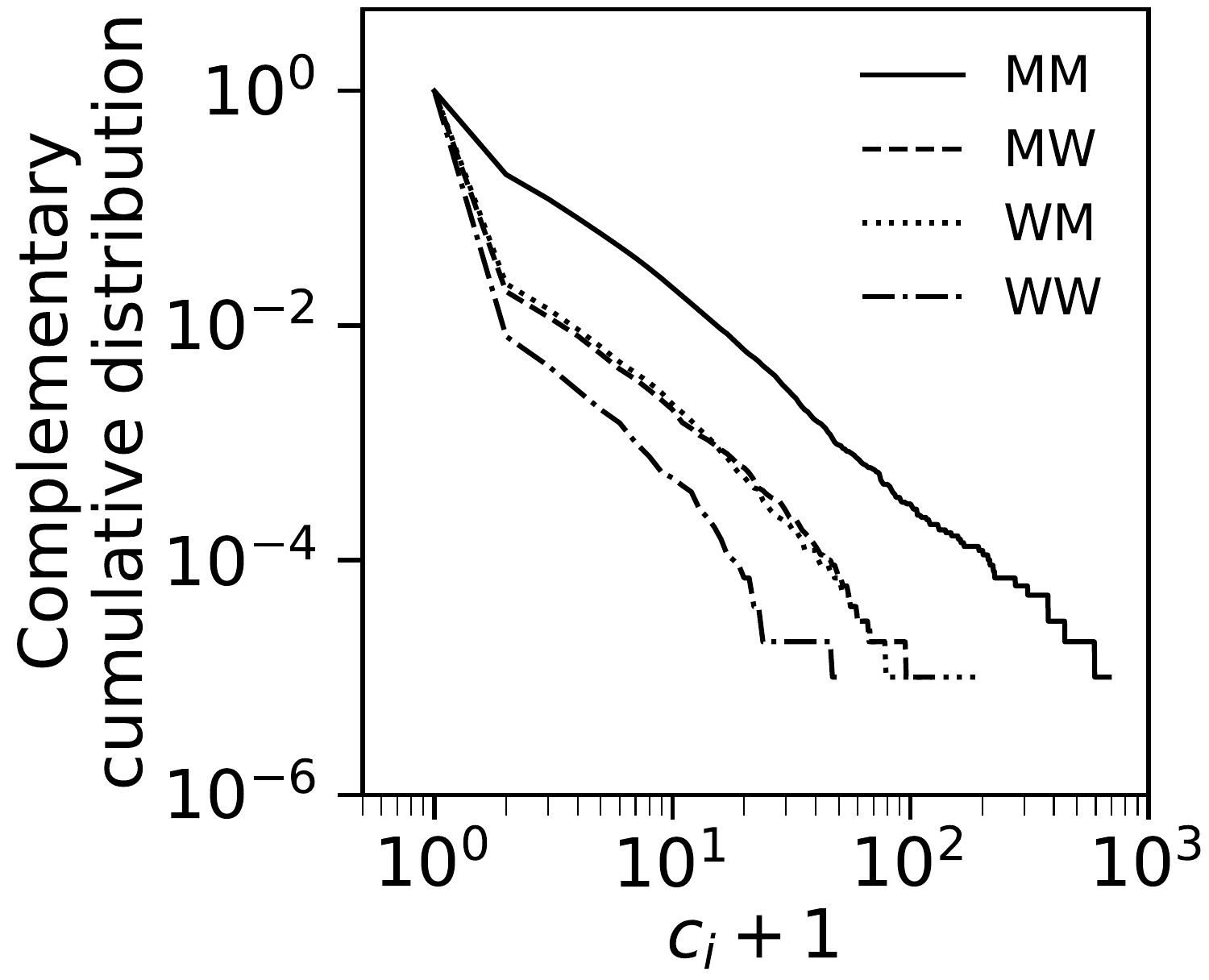}
\caption{
Survival function of the number of citations received by the paper in each gender category.
}
\label{fig:6}
\end{figure}

Tables \ref{table:3} and \ref{table:4} show additional statistics for our dataset.
Figure \ref{fig:6} shows the survival function of the number of citations received by the paper in each gender category.

\section{Appendix B: Pseudocodes}

Algorithms \ref{alg:1}--\ref{alg:3} show the pseudocodes of the three reference models.

\section{Appendix C: Bootstrapping}

We perform bootstrap resampling \cite{dworkin2020, teich2022} to estimate confidence intervals for the over/under-citations shown in Figs.~\ref{fig:2}--\ref{fig:4}.
First, we sample $N$ papers with replacement from the set $V$ uniformly at random.
Then, based on the citations with replacement made by these $N$ sampled papers, we compute the over/under-citation for a given combination of gender category $g$ and reference model $r$.
We repeat this resampling process 500 times independently.
Based on these resamples, we estimate 95\% confidence intervals for the over/under-citation.
Tables \ref{table:5}, \ref{table:6}, and \ref{table:7} show estimates of the confidence intervals for the over/under-citations shown in Figs.~\ref{fig:2}, \ref{fig:3}, and \ref{fig:4}, respectively.

\begin{table}[h]
 \begin{center}
   \caption{Number of papers by conference rank and gender category.}
   \label{table:3}
\begin{tabular}{|l|c|c|c|c|} \hline 
   Rank & MM & MW & WM & WW \\ \hline \hline
   $\text{A}^*$ & 23,258 & 2,188 & 2,602 & 825 \\ 
   A & 19,816 & 2,178 & 2,696 & 1,135\\ 
   B & 22,792 & 2,616 & 3,301 & 1,145 \\
   C & 10,696 & 1,299 & 1,880 & 902\\ \hline
 \end{tabular}
 \end{center}
\end{table}

\begin{algorithm}[h]
\caption{Random-draws model.}
\label{alg:1}
\begin{algorithmic}[1]
\REQUIRE Citation network: $(V, E)$.
\ENSURE $(\bar{w}_{ij, \text{RD}})_{1 \leq i \leq N, 1 \leq j \leq N}, (\bar{c}_{j, \text{RD}})_{1 \leq j \leq N}$
\STATE{Initialize $\bar{w}_{ij, \text{RD}}$ with zero for any $i=1, \ldots, N$ and any $j=1, \ldots, N$.}
\STATE{Initialize $\bar{c}_{j, \text{RD}}$ with zero for any $j=1, \ldots, N$.}
\FOR{$i=1, \ldots, N$}
\STATE{Compute the set $\overline{V}_{\text{RD}}(i)$.}
\FOR{each $v_j \in \overline{V}_{\text{RD}}(i)$}
\STATE{$\bar{w}_{ij, \text{RD}} \leftarrow k_i / |\overline{V}_{\text{RD}}(i)|$.}
\STATE{$\bar{c}_{j, \text{RD}} \leftarrow \bar{c}_{j, \text{RD}} + \bar{w}_{ij, \text{RD}}$.}
\ENDFOR
\ENDFOR
\RETURN{$(\bar{w}_{ij, \text{RD}})_{1 \leq i \leq N, 1 \leq j \leq N}, (\bar{c}_{j, \text{RD}})_{1 \leq j \leq N}$}
\end{algorithmic}
\end{algorithm}

\begin{algorithm}[h]
\caption{Homophilic-draws model.}
\label{alg:2}
\begin{algorithmic}[1]
\REQUIRE Citation network: $(V, E)$ and set of properties: $S$.
\ENSURE $(\bar{w}_{ij, \text{HD}})_{1 \leq i \leq N, 1 \leq j \leq N}, (\bar{c}_{j, \text{HD}})_{1 \leq j \leq N}$
\STATE{Initialize $\bar{w}_{ij, \text{HD}}$ with zero for any $i=1, \ldots, N$ and any $j=1, \ldots, N$.}
\STATE{Initialize $\bar{c}_{j, \text{HD}}$ with zero for any $j=1, \ldots, N$.}
\FOR{each citation $(v_i, v_{i'}) \in E$}
\STATE{Compute the set $\overline{V}_{\text{HD}}(i, i', S)$.}
\FOR{each $v_j \in \overline{V}_{\text{HD}}(i, i', S)$}
\STATE{$\bar{w}_{ij, \text{HD}} \leftarrow \bar{w}_{ij, \text{HD}} + 1 / |\overline{V}_{\text{HD}}(i, i', S)|$.}
\STATE{$\bar{c}_{j, \text{HD}} \leftarrow \bar{c}_{j, \text{HD}} + 1 / |\overline{V}_{\text{HD}}(i, i', S)|$.}
\ENDFOR
\ENDFOR
\RETURN{$(\bar{w}_{ij, \text{HD}})_{1 \leq i \leq N, 1 \leq j \leq N}, (\bar{c}_{j, \text{HD}})_{1 \leq j \leq N}$}
\end{algorithmic}
\end{algorithm}

\begin{algorithm}[h]
\caption{Preferential-draws model.}
\label{alg:3}
\begin{algorithmic}[1]
\REQUIRE Citation network: $(V, E)$ and set of properties: $S$.
\ENSURE $(\bar{w}_{ij, \text{PD}})_{1 \leq i \leq N, 1 \leq j \leq N}, (\bar{c}_{j, \text{PD}})_{1 \leq j \leq N}$
\STATE{Initialize $\bar{w}_{ij, \text{PD}}$ with zero for any $i=1, \ldots, N$ and any $j=1, \ldots, N$.}
\STATE{Initialize $\bar{c}_{j, l, \text{PD}}$ with zero for any $j=1, \ldots, N$ for any $l=1, \ldots, N$.}
\STATE{Sort the papers by publication date in ascending order: $v_{x_1}, \ldots, v_{x_N}$.}
\FOR{$l=2, \ldots, N$}
\FOR{each citation $(v_{x_l}, v_{i'}) \in E$}
\STATE{Compute the set $\overline{V}_{\text{HD}}(x_l, i', S)$.}
\STATE{$\overline{V}_{\text{PD}}(l, i', S) \leftarrow \{v_j\ |\ v_j \in \overline{V}_{\text{HD}}(x_l, i', S) \ \land \ \bar{c}_{j, l-1, \text{PD}} = \bar{c}_{i', l-1, \text{PD}}\}$.}
\FOR{each $v_j \in \overline{V}_{\text{PD}}(l, i', S)$}
\STATE{$\bar{w}_{ij, \text{PD}} \leftarrow \bar{w}_{ij, \text{PD}} + 1 / |\overline{V}_{\text{PD}}(l, i', S)|$.}
\FOR{$z=l, \ldots, N$}
\STATE{$\bar{c}_{j, z, \text{PD}} \leftarrow \bar{c}_{j, z, \text{PD}} + 1 / |\overline{V}_{\text{PD}}(l, i', S)|$.}
\ENDFOR
\ENDFOR
\ENDFOR
\ENDFOR
\FOR{$j=1, \ldots, N$}
\STATE{$\bar{c}_{j, \text{PD}} \leftarrow \bar{c}_{j, N, \text{PD}}$.}
\ENDFOR
\RETURN{$(\bar{w}_{ij, \text{PD}})_{1 \leq i \leq N, 1 \leq j \leq N}, (\bar{c}_{j, \text{PD}})_{1 \leq j \leq N}$}
\end{algorithmic}
\end{algorithm}

\begin{table*}[t]
 \begin{center}
   \caption{Number of papers by subfield and gender category.}
   \label{table:4}
\begin{tabular}{|l|c|c|c|c|} \hline
   Subfield & MM & MW & WM & WW \\ \hline \hline
   Artificial Intelligence & 18,041 & 2,110 & 2,434 & 1,050 \\ \hline
   Computational Theory and Mathematics & 2,483 & 184 & 230 & 91 \\ \hline
   Computer Graphics and Computer-Aided Design& 386 & 23 & 21 & 12 \\ \hline
   Computer Networks and Communications & 11,067 & 1,056 & 1,304 & 267 \\ \hline
   Computer Science Applications & 1,099 & 252 & 330 & 259 \\ \hline
   Computer Vision and Pattern Recognition & 8,853 & 777 & 1,017 & 247 \\ \hline
   Hardware and Architecture & 4,017 & 287 & 310 & 64 \\ \hline
   Human-Computer Interaction & 1,284 & 250 & 291 & 142 \\ \hline
   Information Systems & 6,020 & 777 & 963 & 402 \\ \hline
   Signal Processing & 2,394 & 232 & 339 & 79 \\ \hline
   Software & 1,206 & 159 & 173 & 82 \\ \hline
 \end{tabular}
 \end{center}
\end{table*}

\begin{table*}[t]
 \begin{center}
   \caption{Estimates of 95\% confidence intervals for the over/under-citation computed for each reference model.}
   \label{table:5}
\begin{tabular}{|c|l|B|B|B|B|} \hline
   Made by & \multicolumn{1}{|c|}{Reference model} & To MM & To MW & To WM & To WW \\ \hline \hline
   \multirow{3}{*}{Any} & Random-draws model & $[3.78, 3.81]$ & $[-5.46, -5.31]$ & $[-12.31, -12.17]$ & $[-31.78, -31.58]$ \\ 
   & Homophilic–draws model & $[2.22, 2.24]$ & $[-3.65, -3.52]$ & $[-7.76, -7.64]$ & $[-21.11, -20.92]$ \\ 
   & Preferential–draws model & $[2.23, 2.25]$ & $[-3.65, -3.51]$ & $[-7.86, -7.74]$ & $[-21.11, -20.93]$ \\ 
   \hline
   \multirow{3}{*}{MM} & Random-draws model & $[5.91, 5.94]$ & $[-11.37, -11.20]$ & $[-18.31, -18.16]$ & $[-45.90, -45.72]$ \\ 
   & Homophilic–draws model & $[2.69, 2.71]$ & $[-6.02, -5.86]$ & $[-9.70, -9.56]$ & $[-27.68, -27.47]$ \\ 
   & Preferential–draws model & $[2.71, 2.73]$ & $[-6.07, -5.91]$ & $[-9.79, -9.66]$ & $[-27.65, -27.43]$ \\ 
   \hline
   \multirow{3}{*}{WW} & Random-draws model & $[-13.65, -13.48]$ & $[28.07, 29.07]$ & $[39.23, 40.13]$ & $[103.84, 105.67]$ \\ 
   & Homophilic–draws model & $[-3.02, -2.88]$ & $[2.34, 2.95]$ & $[7.77, 8.32]$ & $[9.29, 10.00]$ \\ 
   & Preferential–draws model & $[-3.10, -2.96]$ & $[2.48, 3.13]$ & $[7.96, 8.53]$ & $[9.47, 10.20]$ \\ 
   \hline
 \end{tabular}
 \end{center}
\end{table*}

\begin{table*}[t]
 \begin{center}
   \caption{Estimates of 95\% confidence intervals for the over/under-citation received by papers in each conference rank.}
   \label{table:6}
\begin{tabular}{|l|B|B|} \hline
   Rank & MM & WW \\ \hline \hline
   $\text{A}^*$ & $[2.60, 2.63]$ & $[-35.24, -34.96]$ \\ 
   A & $[2.05, 2.09]$ & $[-11.14, -10.79]$ \\ 
   B & $[1.30, 1.36]$ & $[-17.28, -16.86]$ \\ 
   C & $[1.68, 1.75]$ & $[-9.39, -8.92]$ \\ 
   \hline
 \end{tabular}
 \end{center}
\end{table*}

\begin{table*}[t]
 \begin{center}
   \caption{Estimates of 95\% confidence intervals for the over/under-citation received by papers in each subfield.}
   \label{table:7}
\begin{tabular}{|l|B|B|} \hline
   Subfield & MM & WW \\ \hline \hline
   Artificial Intelligence & $[4.63, 4.67]$ & $[-26.05, -25.72]$ \\ \hline
   Computational Theory and Mathematics & $[-0.05, 0.05]$ & $[1.50, 2.99]$ \\ \hline
   Computer Graphics and Computer-Aided Design & $[0.35, 0.56]$ & $[-37.15, -32.75]$ \\ \hline
   Computer Networks and Communications & $[0.10, 0.16]$ & $[-32.00, -31.23]$ \\ \hline
   Computer Science Applications & $[-0.98, -0.73]$ & $[-20.77, -19.99]$ \\ \hline
   Computer Vision and Pattern Recognition & $[3.57, 3.61]$ & $[-27.03, -26.38]$ \\ \hline
   Hardware and Architecture & $[0.38, 0.45]$ & $[-22.25, -20.86]$ \\ \hline
   Human-Computer Interaction & $[3.41, 3.59]$ & $[-10.89, -9.85]$ \\ \hline
   Information Systems & $[1.77, 1.85]$ & $[-34.58, -34.03]$  \\ \hline
   Signal Processing & $[-0.12, 0.01]$ & $[13.05, 14.81]$ \\ \hline
   Software & $[-0.33, -0.14]$ & $[-48.87, -47.95]$ \\ \hline
 \end{tabular}
 \end{center}
\end{table*}

\end{document}